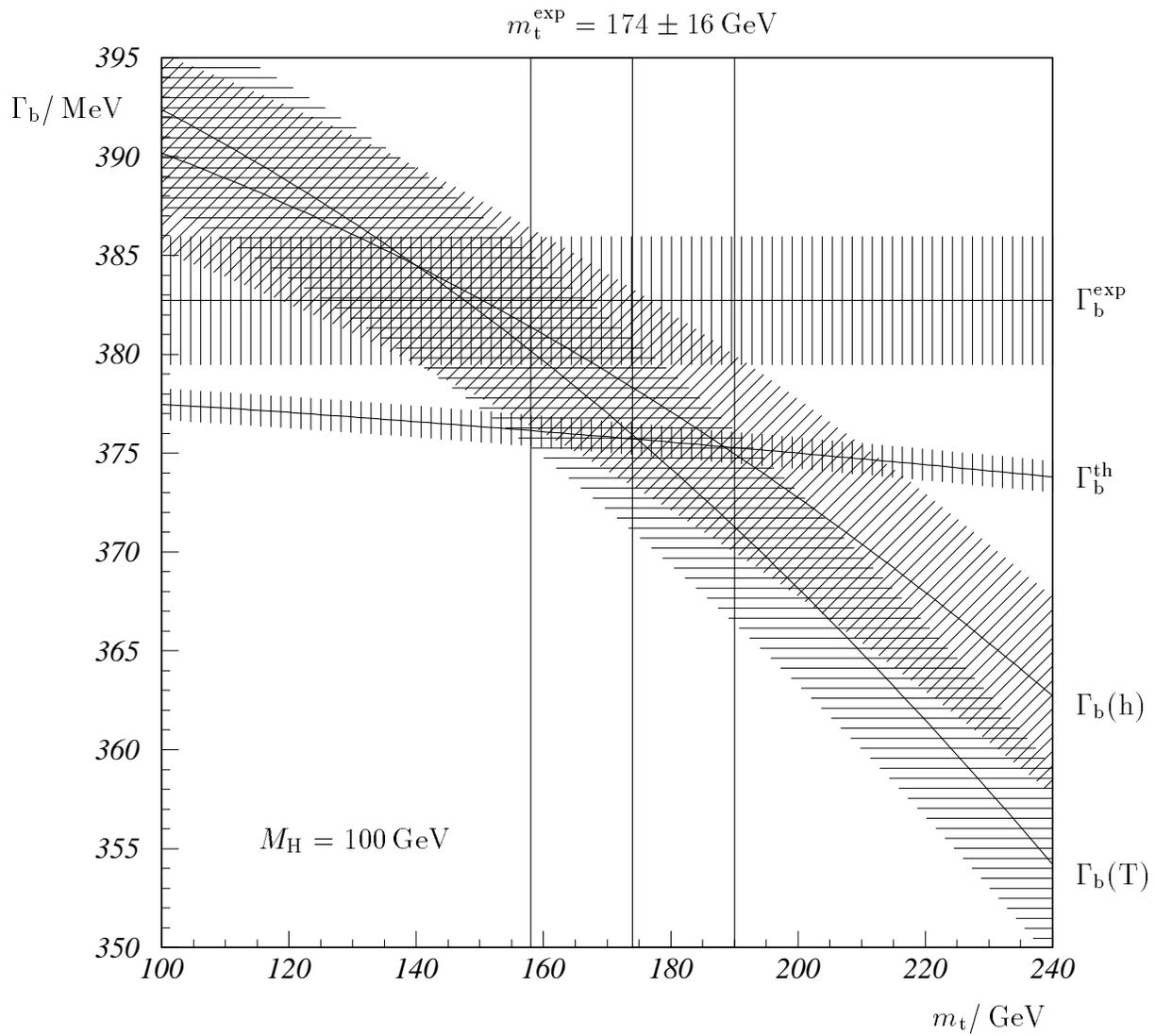

Figure 1:

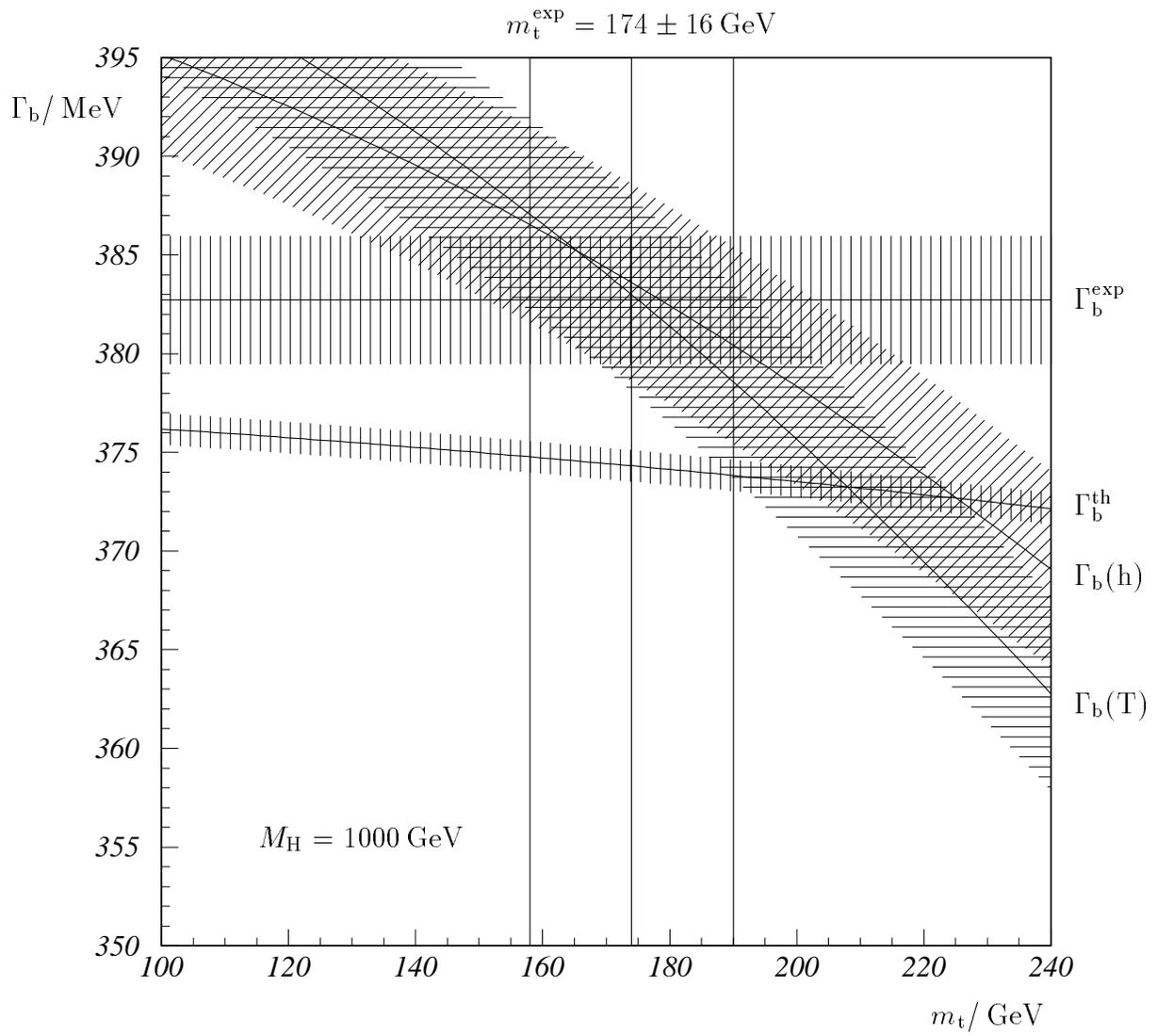

Figure 2:

# A Remark on the $Z^0 \to b\bar{b}$ Width

D. Schildknecht

Department of Theoretical Physics
University of Bielefeld

**Abstract**

The $Z^0 \to b\bar{b}$ width, $\Gamma_b$, is analysed in conjunction with the total and hadronic $Z^0$ widths, $\Gamma_T$ and $\Gamma_h$. Assuming, tentatively, that the present $2\sigma$ discrepancy in $\Gamma_b$ will substantiate as time goes on, for large values of $m_H$ it will be sufficient to modify the $Z^0 b\bar{b}$ vertex only. In contrast, for small values of $m_H$, the theoretical predictions for both the $Z^0$ width into light quarks and leptons as well as the $Z^0 \to b\bar{b}$ vertex will have to be modified.

The precise agreement (e.g. ref. [1]) between the predictions of the $SU(2)_L \times U(1)_Y$ electroweak theory [2] and the experimental data [3] is remarkable indeed. The only evidence for a possible discrepancy between theory and experiment was found in the value of the $Z^0 \to b\bar{b}$ width, which deviates from the theoretical prediction by approximately two standard deviations. The data are consistent with the width predicted for $Z^0 \to d\bar{d}$, and accordingly, they do not show the effect expected from the presence of the mass of the heavy top quark in the $Z^0 b\bar{b}$ vertex. As the discrepancy amounts to two standard deviations only, it may be wise to wait for further analysis of forthcoming data before reflecting too much on a possible theoretical explanation of it.

In the present note, nevertheless, we deal with the $Z^0 \to b\bar{b}$ width, restricting ourselves, however, to a few general comments on how the $Z^0 \to b\bar{b}$ "anomaly" could be accommodated in case it will substantiate and stand the test of time. We will biefly analyse the data on $\Gamma_b$ in conjunction with the data on the total and hadronic $Z^0$ widths, $\Gamma_T$ and $\Gamma_h$, respectively, in comparison with standard predictions. Our essential point consists of the observation that low and high values of the Higgs mass $m_H$, require different dominant modifications of the theory in order to accommodate the experimental value of $\Gamma_b$ in conjunction with the experimental data for $\Gamma_T$ and $\Gamma_h$.

Our analysis will be based on the experimental data presented at the Glasgow Conference [3],

$$M_Z = 91.1888 \pm 0.0044 GeV,$$
$$\Gamma_T = 2497.4 \pm 3.8 MeV,$$
$$R = \Gamma_h/\Gamma_l = 20.795 \pm 0.040, \quad (1)$$
$$\sigma_h = \frac{12\pi \Gamma_l \Gamma_h}{M_Z^2 \Gamma_T^2} = 41.49 \pm 0.12 nb.$$

From the values of $R$ and $\sigma_h$ one derives [1] *

$$\Gamma_l = 83.96 \pm 0.18 \ MeV,$$
$$\Gamma_h = 1746 \pm 4 \ MeV, \quad (2)$$

---

* The correlation matrix between $\Gamma_T, R$ and $\sigma_h$ was taken into account.

and from the measured value of **

$$R_{bh} = \frac{\Gamma_b}{\Gamma_h} = 0.2192 \pm 0.0018, \tag{3}$$

one then obtains

$$\Gamma_b = 382.7 \pm 3.3 \ MeV, \tag{4}$$

In what follows, we will compare the data for $\Gamma_b$ in conjunction with the ones for $\Gamma_T$ and $\Gamma_h$ with standard theoretical predictions. All three of these quantities can be simultaneously analysed in a unified manner by first of all extracting the $Z^0 \to b\bar{b}$ width from the experimental total and hadronic widths, $\Gamma_T^{exp}$ and $\Gamma_h^{exp}$, respectively, via

$$\Gamma_b(T) \equiv \Gamma_T^{exp} - 2\left(\Gamma_u^{th} + \Gamma_d^{th}\right) - 3\left(\Gamma_e^{th} + \Gamma_\nu^{th}\right) \tag{5}$$

and

$$\Gamma_b(h) \equiv \Gamma_h^{exp} - 2\left(\Gamma_u^{th} + \Gamma_d^{th}\right). \tag{6}$$

In these formulae, $\Gamma_u^{th}, \Gamma_d^{th}$, etc. denote the (radiatively corrected) theoretical partial $Z^0$ widths for the $Z^0 \to u\bar{u}$, $Z^0 \to d\bar{d}$, etc. decays, while $\Gamma_b(T)$ and $\Gamma_b(h)$ refer to the partial widths for the $Z^0 \to b\bar{b}$ decay extracted from the total and hadronic $Z^0$ widths, $\Gamma_T$ and $\Gamma_h$, respectively. It is evident that $\Gamma_b(T)$ and $\Gamma_b(h)$ in (5), (6), are "semi-experimental" quantities. They depend on the experimental data on the total and hadronic $Z^0$ widths, $\Gamma_T^{exp}$ and $\Gamma_h^{exp}$, as well as the theoretical predictions for the other partial $Z^0$ widths which are subtracted on the right-hand-sides in (5), (6). Due to the strong dependence on the mass of the top quark, $m_t$ (via the leading $m_t^2$ dependence), also $\Gamma_b(T)$ and $\Gamma_b(h)$ will be decreasing functions of $m_t$. In addition, $\Gamma_b(T)$ and $\Gamma_b(h)$ will depend on the Higgs mass, $m_H$, via $\ln m_H$.

Upon inserting the necessary theoretical partial widths into (5) and (6), we will compare $\Gamma_b(T)$ and $\Gamma_b(h)$ with the theoretical prediction for the $Z^0 \to b\bar{b}$ width, $\Gamma_b^{th}$, and with the experimental one, $\Gamma_b^{exp}$, and draw our conclusions.

---

** This value of $R_{bh}$ is obtained [3] upon fixing $R_c \equiv \Gamma_c/\Gamma_h$ to its Standard Model value of $R_c = 0.171$.

The theoretical values for partial decay widths of the $Z^0$ into leptons and quarks are taken from our recent analysis of the electroweak precision data [1], based on

$$\alpha\left(M_Z^2\right)^{-1} = 128.87 \pm 0.12, \tag{7}$$
$$G_\mu = 1.16639\,(2) \cdot 10^{-5} GeV$$

as well as $M_Z$ from (1) and

$$\alpha_s = 0.118 \pm 0.007, \tag{8}$$
$$m_b = 4.5 GeV$$

as input parameters.

The results of the present analysis are presented in figs. 1,2 for the two cases of a low value of $m_H = 100 GeV$ and a high value of $m_H = 1000 GeV$, respectively.

We first of all consider the case of $m_H = 100 GeV$ shown in fig. 1. From this figure one finds rough agreement of the $Z^0 \to b\bar{b}$ width extracted from the total and hadronic widths with the theoretical prediction, $\Gamma_b^{th}$, i.e.

$$\Gamma_b(T) \cong \Gamma_b(h) \cong \Gamma_b^{th} \tag{9}$$

for

$$m_t \cong 175\ GeV, \tag{10}$$
$$m_H \cong 100\ GeV.$$

Obviously, the result (9), (10) is nothing else but the (known) consistency between theory and experiment in the total $Z^0$ width and in the hadronic $Z^0$ width, expressed, however, in terms of the $Z^0 \to b\bar{b}$ partial width. This consistency holds for values of $m_t \cong 175\ GeV$, the value favored by the results of the direct searches for the top quark [4.]. To remove the (indication of a small) discrepancy with $\Gamma_b^{exp}$ in fig. 1, both, the theoretical prediction for $Z^0 \to b\bar{b}$ decay, $\Gamma_b^{th}$, as well as $\Gamma_b(T)$ and $\Gamma_b(h)$ will have to be modified, in order to keep the validity of (9). According to (5) and (6), this implies that the theoretical predictions for the $Z^0$ widths into light leptons and quarks will have to decrease. In summary, for small values of $m_H$, the data — always assuming that the minor discrepancy between theory and experiment visible at present will substantiate — require a modification of the theory which enlarges $\Gamma_b^{th}$ and diminishes $\Gamma_u^{th}, \Gamma_d^{th}$, etc.

The situation (for $m_t \cong 175\ GeV$) is different in the case of the other extreme, a large mass of the Higgs boson of e.g. $m_H = 1000\ GeV$, as shown in fig. 2. In contrast to (9)

we now have

$$\Gamma_b(T) \cong \Gamma_b(h) \cong \Gamma_b^{exp} \qquad (11)$$

for

$$\begin{aligned} m_t &\cong 175 \ GeV, \\ m_H &\cong 1000 \ GeV. \end{aligned} \qquad (12)$$

For large values of $m_H$ the (theoretical) values for the $Z^0$ widths into light quarks and leptons in (5), (6) are sufficiently suppressed to accommodate the present enhanced experimental value of $\Gamma_b^{exp}$ within the total and hadronic widths, $\Gamma_T^{exp}$ and $\Gamma_h^{exp}$. Accordingly, in this case, it will be sufficient to modify the $Z^0 b\bar{b}$ vertex to obtain consistency with the data for $\Gamma_b^{exp}$ as well as $\Gamma_T^{exp}$ and $\Gamma_h^{exp}$.

In conclusion, the presentation of the data given in figs. 1, 2 clearly illustrates the delicate interplay of the different experimental results and the parameters $m_t$ and $m_H$. If the $2\sigma$ effect in $\Gamma_b$ will stand the test of time, its theoretical explanation will have to discriminate between the low-$m_H$ and the high-$m_H$ options (always assuming $m_t \cong 175 \ GeV$). For low values of $m_H$ the theoretical predictions for the $Z^0$ widths into the light quarks and leptons as well as the $Z^0 \to b\bar{b}$ width will have to be modified. On the other hand, in the limit of large values of $m_H$, it will dominantly only be the theoretical prediction for the $Z^0 \to b\bar{b}$ vertex which must be changed.

### Acknowledgement

The author would like to thank Stefan Dittmaier for fruitful collaboration on electroweak interactions and help in the presentation of the results in figures 1, 2.

# References

[1] S. Dittmaier, D. Schildknecht, M. Kuroda, Bielefeld-preprint
    **BI-TP 94/62**, hep-ph/9501404.

[2] S.L. Glashow, Nucl.Phys.B **22** (1961) 579;
    S. Weinberg, Phys.Rev.Lett. **19** (1967) 1264;
    A. Salam, in: Elementary Particle Theory ed. N. Svartholm (Almquist and Wiksell, 1968), p. 367.

[3] D. Schaile, plenary talk given at the *27th International Conference of High Energy Physics*, Glasgow, July 1994,
    LEP collaborations, preprint CERN/PPE/94-187.

[4] F. Abe et al., CDF Collaboration, Phys.Rev. **D50** (1995) 2966.

**Fig. 1:**

In addition to $\Gamma_b^{exp}$, the figure shows $\Gamma_b^{th}$ as a function of the mass of the top quark, $m_t$, as well as the "semi-experimental" quantities $\Gamma_b(T)$ and $\Gamma_b(h)$ obtained from the total and hadronic $Z^0$ widths, $\Gamma_T$ and $\Gamma_h$, by subtracting the theoretical predictions for the $Z^0$ decay widths into light quarks and leptons. The value of $m_t = 174 \pm 16 \ GeV$ preferred by the CDF searches is also indicated. For the theoretical prediction for $\Gamma_b^{th}$ and for $\Gamma_b(T)$ and $\Gamma_b(h)$ a Higgs-boson of mass of $m_H = 100 \ GeV$ was adopted. The error in $\Gamma_b^{th}$ is due to the experimental error in $\alpha_s$. This error is also taken into account in $\Gamma_b(T)$ and $\Gamma_b(h)$.

**Fig 2.:**

As fig 1, but for $m_H = 1000 \ GeV$.